# From antiferromagnetism to superconductivity in $Fe_{1+y}(Te_{1-x},Se_x)$ ($0 \leq x \leq 0.20$): a neutron powder diffraction analysis

A. Martinelli[1,*], A. Palenzona[1,2], M. Tropeano[1,3], C. Ferdeghini[1], M. Putti[1,3], M. R. Cimberle[4], T. D. Nguyen[5], M. Affronte[5], C. Ritter[6]

[1]*CNR-INFM-LAMIA Corso Perrone 24, 16152 Genova – Italy*

[2]*Dipartimento di Chimica e Chimica Industriale, Università di Genova, via Dodecaneso 31, 16146 Genova – Italy*

[3]*Dipartimento di Fisica, Università di Genova, via Dodecaneso 33, 16146 Genova – Italy*

[4]*CNR-IMEM, Dipartimento di Fisica, Via Dodecaneso 33, 16146 Genova, Italy*

[5]*CNR-INFM-S3 and Dipartimento di Fisica Università di Modena e Reggio Emilia, via G. Campi 213, 41125 Modena, Italy*

[6]*Institute Laue - Langevin, 6 rue Jules Horowitz, 38042 Grenoble Cedex 9 – France*

**Abstract**

The nuclear and magnetic structure of $Fe_{1+y}(Te_{1-x},Se_x)$ ($0 \leq x \leq 0.20$) compounds was analyzed between 2 K and 300 K by means of Rietveld refinement of neutron powder diffraction data. Samples with $x \leq 0.075$ undergo a tetragonal to monoclinic phase transition at low temperature, whose critical temperature decreases with increasing Se content; this structural transition is strictly coupled to a long range antiferromagnetic ordering at the Fe site. Both the transition to a monoclinic phase and the long range antiferromagnetism are suppressed for $0.10 \leq x \leq 0.20$. The onset of the structural and of the magnetic transition remains coincident with the increase of Se substitution. The low temperature monoclinic crystal structure has been revised. Superconductivity arises for $x \geq$

[*] Corresponding author: amartin@chimica.unige.it

0.05, therefore a significant region where superconductivity and long range antiferromagnetism coexist is present in the pseudo-binary FeTe - FeSe phase diagram.

**1. Introduction**

The end-members of the solid solution $Fe_{1+y}(Te_{1-x},Se_x)$ belong to the class of compounds characterized by the presence of edge-sharing Fe-centred tetrahedra, forming layered structures that attracted much attention on account of the discovery of a relatively high superconducting transition temperature ($T_c$) in $LaFeAs(O_{1-x}F_x)$.[1] Both FeTe and FeSe are isostructural with $\alpha$-PbO, crystallizing at room temperature in the tetragonal system (*P4/nmm* space group), and they are both referred to as $\beta$ phase.[2,3] It is worth pointing out that $Fe_{1+y}(Te_{1-x},Se_x)$ compounds are often referred to as $\alpha$ phase in the literature, probably because they are isostructural with $\alpha$-PbO. However, in the Fe-Te and Fe-Se phase diagrams the notation $\alpha$ refers only to the terminal Fe-rich solid solutions or to the low temperature monoclinic phase $\alpha$-$Fe_7Se_8$.

Subtle stoichiometric variations in $Fe_{1+y}Se$ can determine different structural and resistive properties: for $y = 0.01$ the phase can exhibit a structural transition on cooling followed by the rise of superconductivity, whereas for $y = 0.03$ the structural change is suppressed as well as the superconductive transition.[4] The structural properties of $Fe_{1+y}Te$ are as well strongly affected by faint stoichiometrical variations: compounds with $y = 0.141$ and $0.076$ crystallize at low temperature in the orthorhombic (*Cmme* space group) and monoclinic ($P2_1/m$ space group) system, respectively.[5] Antiferromagnetic (AFM) spin ordering occurs in these $Fe_{1+y}Te$ compounds at the Fe site at low temperature, characterized by a commensurate magnetic wave vector $\mathbf{q} = (½\ 0\ ½)$ when it is associated to the monoclinic structure and incommensurate in the orthorhombic one.[5] Partial substitution of Te with Se progressively suppresses the tetragonal to monoclinic phase transition and induces superconductivity at low temperature. Static long-range (LR) magnetic order is suppressed as well on Se doping, although short-range (SR) not commensurate AFM fluctuations are still present in samples with $x = 0.25$ exhibiting a trace of superconductivity,[6] as well as in superconducting samples with $x$ ranging from 0.30 to 0.416.[5,7] Coexistence between incommensurate SR-AFM and superconductivity has been also reported by Khasanov *et al.* [8] in samples with $0.25 \leq x \leq 0.40$, whereas commensurate magnetic order prevails for $x \leq 0.10$.

At present it is not yet clear whether the tetragonal to monoclinic transition occurring in FeTe on cooling is driven by magnetism or not; Li *et al.* [7] argue from specific heat measurements that this structural transition is driven by magnetism. Conversely McQueen *et al.* [4] analyzed the tetragonal to orthorhombic transition taking place in FeSe. As there is no magnetic order present even at low temperature in this compound they concluded that magnetism is not the driving force for the structural transition even in the other superconducting pnictides.

By means of neutron powder diffraction we analyze in this paper the evolution from the AFM ground state of FeTe to the superconducting one and the suppression of the structural and magnetic transition by substituting Te with Se up to $x = 0.20$.

## 2. Experimental

Samples of $Fe_{1+y}(Te_{1-x},Se_x)$ with $x$ = 0.00, 0.025, 0.05, 0.075, 0.10, 0.15, 0.20 were prepared by means of a solid state reaction in two steps: 1) a mixture of stoichiometric amounts of pure elements was reacted in a sealed evacuated Pyrex tube at 400-450°C for 15-20 h; 2) the so-obtained product was ground, pelletised and heated at 800°C for 7-8 days in a sealed evacuated quartz tube. All the operations were carried out in a glove box with $O_2$ and $H_2O$ partial pressures less than 1 ppm.

Neutron powder diffraction (NPD) experiments were carried out at the Institute Laue Langevin (Grenoble – France). In order to evaluate the temperature at which the structural and magnetic transitions take place, thermo-diffractograms of the samples with $x$ = 0.00, 0.025, 0.05, 0.075 were acquired on heating in continuous scanning mode in the $T$ range 1.5 – 80 K using the high intensity D1B diffractometer ($\lambda$ = 2.52 Å). High resolution NPD patterns were collected at selected temperatures between 2 and 300 K using the D1A diffractometer ($\lambda$ = 1.91 Å) for the samples with $x$ = 0.00, 0.05, 0.10, 0.15, 0.20. Rietveld refinement of NPD data was carried out using the program

FULLPROF;[9] by means of a NAC standard an instrumental resolution file was obtained and applied during refinements in order to detect micro-structural contributions to the NPD peak shape. The diffraction lines were modelled by a Thompson-Cox-Hastings pseudo-Voigt peak shape function convoluted with an axial divergence asymmetry function. The background was fitted by a fifth-order polynomial. The following parameters were refined in the last cycle of calculation: the overall scale factor; the background (five parameters of the $5^{th}$ order polynomial); $2q$-Zero; the unit cell parameters; the specimen displacement; the reflection-profile asymmetry; the Wyckoff positions not constrained by symmetry; the isotropic thermal parameters $B$; the anisotropic strain parameters.

Resistive measurements were carried out in a Physical Properties Measurements System (PPMS, Quantum Design) in the temperature range 5-300 K.

Heat capacity was measured on pellets by a Quantum Design PPM System. The heat pulse was fixed in order to have 1% in temperature response and the two-tau relaxation method was necessary to account for the low thermal conductivity of the samples.

### 3. Results and discussion

*3.1 Resistivity*

Resistivity measurements of $Fe_{1+y}(Te_{1-x},Se_x)$ samples are shown in Figure 1. With decreasing temperature the resistivity of all the samples exhibit a logarithmic upturn which progressively flattens with increasing $x$ (see ref.[10] for a more detailed analysis). At $T \sim 77$ K, FeTe shows a steep drop in the resistivity and the behaviour become quite metallic below $T \sim 70$ K. Samples with up to 7.5% of Se content, the effect of Se substitution results in a broadening and a shifting to lower temperature of the drop. Samples with $x \geq 0.1$ do not show any feature related to this drop but a flattening of the resistivity followed by an upturn just before the superconducting transition. The superconductive critical temperatures ($T_c$) increase with Se substitution (Table I), from 11.0 K for the $x = 0.05$ sample up to 13.6 K for the $x = 0.2$ sample.

*3.2 Specific heat*

Figure 2 shows the temperature dependence of the heat capacity C(*T*) of FeTe and Fe(Te$_{0.975}$Se$_{0.025}$). Slow cooling and heating rates were used and data collected for each thermal cycle. The absolute value of C(*T*) is very similar for both samples as expected for the minor difference introduced by the very low Se substitution. At high temperatures the absolute value of C(*T*) tends to saturate at the value of 6*R* predicted by the Dulong-Petit law. Anomalies are clearly visible around 70 K and 58 K for FeTe and Fe(Te$_{0.975}$Se$_{0.025}$), respectively. Consistent with first order character of this transition thermal hysteresis is present at these transitions (inset of Figure 2) as already reported by Li *et al.*[7]

*3.3 Neutron powder diffraction*

In order to investigate whether the structural transition from *P*4/*nmm* to *P*2$_1$/*m* occurs in FeTe in one step or if it proceeds via an intermediate *Cmme* phase, the evolution of the 101, 111, 112, 200 nuclear peaks was analyzed as a function of temperature using the thermodiffractograms collected on D1B. In fact, all these peaks see a splitting during a tetragonal to monoclinic phase transition, whereas the presence of an intermediate orthorhombic phase would be seen through the splitting of solely the 111 and 112 peaks In order to verify whether peak splitting occurs at different temperatures, the temperature dependence of the Lorentzian isotropic strain (LIS) contribution to the four afore mentioned diffraction peaks was analyzed carrying out a peak fit procedure using an instrumental resolution file of the used diffractometer. This is the most sensitive way of searching for a possible peak broadening caused by an underlying line splitting in cases where the resolution of the diffractometer is not sufficient to detect the splitting directly. Figure 3 shows as an example the evolution of the LIS of the 4 Bragg peaks with temperature (normalised data) as measured for the FeTe sample. A similar behaviour of the same Bragg peaks as a function of temperature was found for all the other measured samples with $x \leq 0.075$, thereby ruling out the occurrence of an intermediate orthorhombic phase. The temperature at which the antiferromagnetic ordering takes place ($T_N$) is revealed by the arising of magnetic peaks in the NPD patterns. Figure 4 (upper panel) shows the evolution of the NPD pattern of FeTe in the regions of the ½0½ magnetic peak and the tetragonal 200 nuclear peak, evidencing the simultaneous occurrence of magnetic ordering ($T_N$) and structural transition ($T_{T-M}$) at about 70 K. A closer analysis of the evolution of the ½0½ magnetic peak intensity and of the LIS of the 200 nuclear peak in this temperature region (Figure 4, lower panel) confirms that both transitions take place simultaneously. $T_{T-M}$ and $T_N$ decrease with increasing Se content, but both transitions remain coupled and coincident. The width of the structural and the magnetic transitions increases with substitution (Figure 5) which can be related to an increasing phase coexistence due to incomplete transition into the monoclinic phase (see below). The important result of this analysis of the temperature dependence of the magnetic and nuclear Bragg peaks is that $T_{T-M} \cong T_N$ whatever the Se content (critical transition temperatures are reported in Table I). This has to be compared to the situation found in other substituted Fe-based superconductors: in $BaFe_2As_2$, belonging to the so-called 122 family of Fe-based superconductors, the first order tetragonal to orthorhombic structural transition and the AFM ordering take place simultaneously; Co substitution leads here to a decrease of both transition temperatures but as well to their separation as the magnetic ordering temperature is more rapidly reduced.[11] In the 1111 family Co substitution induces a more complicated behaviour: in CaFeAsF, where the structural transition takes place at higher $T$ than the magnetic, 5% of Co doping lowers both transition temperatures which become almost coincident.[12,13] Conversely in $La(Fe_{1-x}Co_x)AsO$ the structural transition is more stable against substitution, whereas magnetism gets rapidly suppressed.[14] If substitution takes place at the rare earth site, such as in $(La_{1-y}Y_y)FeAsO$, both transition temperatures are similarly affected by substitution.[15]

Figure 6 (on the left) shows the comparison of the NPD patterns of the four samples in the regions of the ½0½ magnetic peak and of the tetragonal 200 nuclear peak: a progressive decrease of the magnetic scattering as well as of the peak splitting is observed with the increase of Se substitution. Structural data obtained by Rietveld refinement of the high resolution NPD data (D1A diffractometer) collected at 2 K and 300 K are reported in Table II; Figure 7 shows the Rietveld refinement of the NPD data of FeTe collected at 2 K, Figure 8 shows the evolution of the cell parameters as a function of $T$ below 300 K for the 5 measured compounds. The substitution of Te by the smaller Se leads to a steady decrease of the cell volume, whereby this effect is mostly caused by the reduction of the cell parameter $c$; this behaviour is related to the fact that the bond length between the tetrahedral Fe and the chalcogen element ($Ch$ = Te, Se) strongly reduces as the Se content increases, whereas the in-plane Fe-Fe bond lengths is almost unaffected (Table II). The occupancies of the Fe sites were refined: while the Wyckoff site $2a$ is always fully occupied, slightly different occupancies are obtained at the interstitial site $2c$. This leads to the refined stoichiometries of $Fe_{1+y}(Te_{1-x}Se_x)$ with $y$ = 0.05, 0.04, 0.03, 0.02, 0.02 for $x$ = 0.00, 0.05, 0.10, 0.15, 0.20, respectively. It becomes evident that the occupation at the site $2c$ (interstitial Fe) decreases with Se substitution. The tetrahedral layer in $Fe_{1+y}(Te_{1-x},Se_x)$ is much less compressed than the homologous FeAs one in $RE$FeAsO ($RE$: rare earth) compounds; in fact considering the cube inscribing the tetrahedron centred by Fe, in both cases it results compressed along the $c_C$ axis, but in the former case the $c_C/a_C$ ratio is ~0.9 whereas in the latter one is ~0.7 (structural data in this case are those reported in ref. [16] for SmFeAsO; $c_C$ and $a_C$ refers to the edge of the pseudo-cube, not to be confused with the cell parameters).

Small amounts of a secondary phase (~ 5%) are present in all the samples: in particular for $x$ = 0.00 the secondary phase is constituted of orthorhombic $FeTe_2$, whereas for the other samples the hexagonal $Fe_{0.67}Te$ phase is present.

At 2 K the samples with $x$ = 0.00 and 0.05 crystallize in the $P2_1/m$ space group (Table III); note that the monoclinic structural model of Table III slightly differs from the one previously reported for FeTe by Li et al.[7] that located Fe(1) at a $2b$ site with coordinates ¾,¼,$z$ and Te as well as Fe(2) at a $2a$ site with coordinates ¼,¼,$z$. This structural model is quite debatable: in fact in the $P2_1/m$ space group the coordinates of the $2a$ and $2b$ sites are 0,0,0, and ½,0,0, respectively,[17] and the two sites are characterized by different site symmetries; conversely in our structural model all atoms are located at the $2e$ site and hence all sites share the same symmetry.

Using the high resolution data it is possible to ascertain the coexistence of the two polymorphic modifications around $T_{T-M}$; in fact at 70 K, slight below $T_{T-M}$, FeTe is *e.g.* constituted of ~ 70% of monoclinic phase, the remaining being tetragonal, whereas in the compound with 5% of Se the amount of monoclinic phase is only ~ 10% at $T_{T-M}$ = 50 K. The occurrence of phase coexistence is a clear indication that the tetragonal to monoclinic structural transition is of the first order, as emphasized by specific heat measurements.

Both samples with $x$ = 0.00 and 0.05 exhibit an AFM spin ordering at Fe sub-lattice characterized by a commensurate magnetic wave vector **q** = (½ 0 ½) and spins ordered along the *b* axis (inset of Figure 7). No component along the *x* and/or *z* axis can be detected, differently from what has been observed for $Fe_{1.068}Te$;[7] the magnetic moment value is decreases slightly with Se substitution from 2.54(2) $m_B$ to 2.08(2) $m_B$. It is interesting to observe that our $Fe_{1.05}Te$ sample is characterized by the highest value of ordered magnetic moment among those reported in literature.[5,7] A comparison with these data indicates that the value of the magnetic moment of Fe at the 2*a* site might be correlated to the Fe occupation at the 2*c* site with a higher magnetic moment on 2*a* for lower occupancies on 2*c*.

The tetragonal to monoclinic phase transition determines the branching of the Fe-Fe bond distances, as evidenced in Figure 9 for FeTe. In the monoclinic field the arrangement of the Fe-Fe bonds gives rise to a stripe-like pattern oriented along the *y* axis, different from that observed at low temperature in orthorhombic *RE*FeAsO (Figure 10). Probably the different spin orderings of the Fe sub-lattices taking place in FeTe and *RE*FeAsO can be ascribed to the different arrangements of the Fe-Fe bond lengths occurring at low temperature that determine different magnetic exchange paths.

Samples with $x$ = 0.10, 0.15, 0.20 retain the tetragonal P4/*nmm* structure in the whole inspected range, no evidence for a selective peak broadening can be detected (structural data at 2 K in Table IV) and the magnetic scattering is completely suppressed. In a previous NPD investigation[18] on $Fe_{1+y}(Te_{1-x},Se_x)$ compounds a pronounced increase of the full width at half maximum (FWHM) of the 200 diffracting peak was observed below 150 K for samples with $0.125 \leq x \leq 0.50$; this behaviour was related to a decrease of the lattice symmetry. The micro-structural analysis by means of the Williamson-Hall method (see below) of our samples with $0.10 \leq x \leq 0.20$ indicates that not only the $h00$ peaks broaden on cooling, but a similar behaviour characterizes *e.g.* also the $hh0$, the $hhh$ peaks and in general all those peaks with a strong component lying in the $ab$ plane. This kind of peak broadening must be therefore related to structural strains taking place on cooling in the $ab$ plane, rather than to a structural transition. This structural strains probably originates the faint negative thermal expansion along the $a$ axis observed below 100 K (Figure 8), a phenomenon observed in previous NPD investigations[18] and in thermal expansion measurements carried out on Fe(Te$_{0.5}$Se$_{0.5}$) single crystal[19].

The evolution of the cell volume as a function of $T$ can be described by:

$$V(T) = ??_S \cdot U(T) + V(0) \qquad \text{Eq. (1)}$$

where $?_S$ is the adiabatic compressibility, *g* the Grüneisen parameter, $U(T)$ the internal energy and $V(0)$ the volume at the lowest temperature. From D1B-NPD data, the experimental dependence of the cell volume as a function of $T$ between 2 and 80 K is available only for the $x = 0.05$ and 0.075 samples. As evaluated by integrating specific heat data, the samples with $x = 0.00$ and 0.025 are characterized by the same value of $U(T)$ (inset of Figure 11); hence we can confidently assume that also the sample with $x = 0.05$ is characterized by a similar value of $U(T)$. For this reason the value of $U(T)$ estimated by specific heat was introduced in Eq. 1 to describe $V(T)$ of the sample with 5% of Se and compare the result with experimental data (Figure 11). A good agreement is obtained below $T_{T-M} \cong T_N$ (~50 K) for $??_S \sim 2.1 \cdot 10^{-10}$ Pa$^{-1}$, a value similar to that reported for single crystal Fe$_{1.06}$Te in the monoclinic phase region.[19] Above $T_{T-M}$, in the tetragonal phase region, the experimental data depart from those estimated by Eq. 1. Around $T_{T-M}$ this behaviour can be due to the presence of latent heat that is not properly accounted in bare heat capacity data; however with increasing $T$ the reduced thermal expansion takes place. Such a behaviour, observed also for Fe$_{1.06}$Te (see Fig.5 in ref [19]), suggests that in the AFM phase also the magnetic excitation contributes as well to the Grüneisen parameter, whereas this contribution disappears above $T_{T-M} \cong T_N$.

Information on the effect of the Se-substitution on the micro-structure of the samples was obtained analyzing the broadening of NPD lines (high resolution - D1A data) by means of the Williamson-Hall plot method.[20] Generally, in the case where size effects are negligible and the strain is isotropic, a straight line passing through all the points and through the origin has to be observed, where the slope provides the lattice strain: the higher the slope the higher the strain. If the broadening is not isotropic, size and strain effects along some crystallographic directions can be obtained by considering different orders of the same reflection. In our case, for each sample, the size contribution is negligible, since a straight line passing through the origin can be traced. Figure 12 shows the superposition of the Williamson-Hall plots obtained by the micro-structural analysis of the data collected at 300 K on D1A; for clarity only selected diffraction peaks are reported and indexed. It is evident that in pure FeTe micro-strain is almost isotropic, since a straight line can be traced passing approximately through all the diffracting peaks. The Se-substitution induces strain along the [001] direction, increasing with increasing Se content, whereas the strain in the *ab* plane is almost unaffected by the degree of substitution. The compression of the tetrahedral layer, measured as the $c_C/a_C$ ratio of the pseudo-cube inscribing the Fe$Ch_4$ tetrahedron, increases with the increase of the Se content and this is reflected in the changes experienced by the *Ch*-Fe(1)-*Ch* bond angles. Two kinds of *Ch*-Fe(1)-*Ch* bond angles are present, the former bisected by the *c* axis, widening with the substitution, the latter that conversely contracts with the increase of the Se content (Table II). The increase of $T_c$ observed in samples with $x \geq 0.05$ is probably related to these structural and micro-structural features. This is in agreement with the findings of Bud'ko *et al.* [19] which argued that in-plane pressure should decrease $T_c$, whereas a pressure along *c* should lead to an increase. Interestingly, in thin films, the behaviour seems to be exactly the opposite, as $T_c$ increases with the decrease of cell parameter *a*, that is with the increase of the compressive strain in the Fe plane.[21] Note that in these thin films the value of the cell parameter *a* is strongly dependent on both the nature of the substrate and the film thickness, but results in any case shorter than the one measured on the bulk sample.

In this context it is interesting to observe that density functional calculations foresee a transition from the observed double stripe AFM spin ordering and ($\pi$,0) spin fluctuations to a single stripe one and ($\pi$,$\pi$) spin fluctuations as the distance of *Ch* from the Fe plane (half edge of the pseudo-cube along *c*) decreases below ~1.72 Å.[22] In particular it has been suggested that superconductivity in Fe chalcogenides may be related to ($\pi$,$\pi$) spin ordering and fluctuations. As can be easily calculated from the data reported in Table II, in the sample bearing 5% of Se, characterized by $T_N$ ~ 50.0±1 K and $T_c$ = 11.0 K, the distance of *Ch* from the Fe plane is ~1.73 Å; in the sample with 10% this distance decreases to ~1.71 Å, LR-AFM ordering is suppressed and only superconductivity takes place ($T_c$ = 11.9 K). This result is thus in very good agreement with density functional calculations foreseeing superconductivity when ($\pi$,$\pi$) spin ordering and fluctuation become dominant.[22] In any case it must be underlined that several neutron magnetic scattering experiments[5,6,7,8] ascertained that even in the SR-AFM regime, appearing for Se contents exceeding the critical value of about 10 at.%, an incommensurate wave vector **q** = (½-$\delta$ 0 ½) is present, similar to one occurring in the LR-AFM regime; this result suggests that the type of spin fluctuation does not change with Se content. Theoretical calculations suggest that the structural transition is partly determined by orbital ordering, involving a cooperative Jahn-Teller distortion, rather than spin ordering; as a result double exchange interactions can take place leading to the observed AFM structure.[23]

In this context it is worth to note that: 1) SR-AFM incommensurate order[5,6,7,8] has been observed in samples with *x* up to ~0.45, inside which the magnetic volume fractions is strongly dominant (generally exceeding 90%),[8] but characterized by a tetragonal structure at low *T*;[5,7] 2) in both SR and LR-AFM regimes the magnetic wave vector is the same, except for its slight incommensurability in the SR-AFM regime; 3) for pure FeTe the $T_N$ obtained by NPD analysis (sensitive only to LR-AFM interactions) and by muon-spin rotation[8] (sensitive also to SR-AFM interactions) is the same, thus suggesting that SR-AFM interactions are absent above the structural transition; 4) the magnetic transition temperatures reported in literature[8] for SR-AFM ordering are in good agreement with those that we obtained for LR-AFM ordering by analysing the neutron thermodiffractograms, as plotted in Figure 13. These facts suggest that LR-AFM order can take place only when the structural transition is not suppressed by Se-substitution in FeTe. Above a critical value of Se content, located in the range $0.075 \leq x \leq 0.10$, the structural transition is suppressed and only SR magnetic interactions can take place, their percolation into a LR-AFM structure being hindered. Conversely the occurrence of the monoclinic structure favours magnetic interactions, as shown in Figure 13 where the samples undergoing this structural transition are characterized by a steep rise of $T_N$. This conclusion is also supported by the fact that magnetism is not the driving force for the tetragonal to orthorhombic phase transition taking place at 90 K in the tetragonal isostructural and superconducting $Fe_{1.01}Se$ compound.[24] In any case further investigations are needed in order to exactly ascertain the role of magnetic interactions in the tetragonal to orthorhombic structural transition of FeTe.

**Conclusions**

Samples with nominal composition $Fe_{1+y}(Te_{1-x},Se_x)$ ($0 \leq x \leq 0.20$) were analyzed by neutron powder diffraction between 2 K and 300 K. Rietveld refinement indicates that samples with $x \leq 0.075$ undergo a first order tetragonal to monoclinic phase transition, whose temperature decreases with increasing Se content. This structural transition is accompanied by a simultaneous antiferromagnetic ordering at the tetrahedral Fe site. Differently from what has been observed in other families of Fe-based superconductors, the structural and the long range magnetic transitions in $Fe_{1+y}(Te_{1-x},Se_x)$ remain coincident after Se substitution. In samples with $x \geq 0.10$ the structural phase transition as well as the long-range magnetic ordering are suppressed. Superconductivity is observed for $x \geq 0.05$, although zero resistivity occurs only for $x \geq 0.10$. Se substitution produces a cell compression, a decrease of the *Ch* height over the Fe plane and an increase of the *Ch*-Fe-*Ch* tetrahedral bond angle. It is likely that the structural transition is not driven by magnetic interactions, but conversely that long-range magnetic order can take place only if the structural transition is not suppressed by Se-substitution in FeTe.

Table I: Critical temperatures for the tetragonal to monoclinic structural change, the long range AFM ordering and the superconductive transition.

| Se content ($x$) | $T_{\text{T-M}}$ (K) | $T_{\text{N}}$ (K) | $T_{\text{c}}$ (K) |
|:---:|:---:|:---:|:---:|
| 0.00 | 72.5±1 | 72.5±1 | / |
| 0.025 | 62.5±1 | 62.5±1 | / |
| 0.05 | 50.0±1 | 50.0±1 | 11.0 |
| 0.075 | 43.0±1 | 43.0±1 | 11.5 |
| 0.10 | / | / | 11.9 |
| 0.15 | / | / | 12.7 |
| 0.20 | / | / | 13.6 |

Table II: Structural data for $Fe_{1+y}(Te_{1-x},Se_x)$ samples obtained by Rietveld refinement of NPD data collected at 300 K ($P4/nmm$ space group; bond lengths and angles refer to tetrahedral Fe).

| | $x = 0.00$ | $x = 0.05$ | $x = 0.10$ | $x = 0.15$ | $x = 0.20$ |
|---|---|---|---|---|---|
| $a$ (Å) | 3.8219(1) | 3.8184(1) | 3.8160(1) | 3.8133(1) | 3.8114(1) |
| $c$ (Å) | 6.2851(1) | 6.2617(1) | 6.2381(1) | 6.2116(1) | 6.1843(1) |
| Fe(1) - 2a | ¾ ¼ 0 | ¾ ¼ 0 | ¾ ¼ 0 | ¾ ¼ 0 | ¾ ¼ 0 |
| Fe(2) – 2c | ¼ ¼ 0.710(4) | ¼ ¼ 0.718(6) | ¼ ¼ 0.720(8) | ¼ ¼ 0.69(1) | ¼ ¼ 0.675(9) |
| $Ch$ – 2c | ¼ ¼ 0.2792(4) | ¼ ¼ 0.2763(4) | ¼ ¼ 0.2746(4) | ¼ ¼ 0.2729(4) | ¼ ¼ 0.2714(4) |
| Fe-$Ch$ bond length (Å) | 2.594(2)×4 | 2.576(2)×4 | 2.564(2)×4 | 2.551(2)×4 | 2.539(2)×4 |
| Fe-Fe bond length (Å) | 2.702(1)×4 | 2.700(1)×4 | 2.698(1)×4 | 2.696(1)×4 | 2.695(1)×4 |
| $Ch$-Fe-$Ch$ bond angle (deg) | 94.88(5)×2 | 95.63(5)×2 | 96.17(5)×2 | 96.72(5)×2 | 97.26(5)×2 |
| | 117.2(1)×4 | 116.8(1)×4 | 116.5(1)×4 | 116.2(1)×4 | 115.9(1)×4 |
| $R_F$ (%) | 3.52 | 3.19 | 2.98 | 3.17 | 3.48 |
| $R_B$ (%) | 4.47 | 4.13 | 4.15 | 4.27 | 4.48 |

Table III: Structural data for $Fe_{1+y}(Te_{1-x},Se_x)$ samples obtained by Rietveld refinement of NPD data collected at 2 K ($P2_1/m$ space group; bond lengths and angles refer to tetrahedral Fe).

|  | $x = 0.00$ | $x = 0.05$ |
|---|---|---|
| $a$ (Å) | 3.8312(1) | 3.8216(1) |
| $b$ (Å) | 3.7830(1) | 3.7893(1) |
| $c$ (Å) | 6.2643(1) | 6.2331(1) |
| $\beta$ (deg) | 89.17 | 89.37 |
| Fe(1) – 2e | 0.7616(1) ¼ 0.0035(4) | 0.7591(1) ¼ 0.0030(4) |
| Fe(2) – 2e | 0.229(1) ¼ 0.714(1) | 0.246(1) ¼ 0.683(1) |
| Ch – 2e | 0.2580(1) ¼ 0.2800(1) | 0.2571(1) ¼ 0.2751(1) |
| Fe-$Ch$ bond length (Å) | 2.5740(3)×2 | 2.5463(3)×2 |
|  | 2.5962(2)×2 | 2.5630(3)×2 |
| Fe-Fe bond length (Å) | 2.7568(2)×2 | 2.7411(2)×2 |
|  | 2.6295(2)×2 | 2.6420(2)×2 |
| $Ch$-Fe-$Ch$ bond angle (deg) | 93.53(1)×1 | 95.02(1)×1 |
|  | 95.76(2)×1 | 96.83(2)×1 |
|  | 115.56(1)×2 | 115.20(1)×2 |
|  | 119.08(1)×2 | 118.03(1)×2 |
| $m(\mu_B)$ | 2.54(2) | 2.08(2) |
| $R_F$ (%) | 3.52 | 2.70 |
| $R_B$ (%) | 4.47 | 3.68 |

Table IV: Structural data for $Fe_{1+y}(Te_{1-x},Se_x)$ samples obtained by Rietveld refinement of NPD data collected at 2 K ($P4/nmm$ space group; bond lengths and angles refer to tetrahedral Fe).

|  | $x = 0.10$ | $x = 0.15$ | $x = 0.20$ |
|---|---|---|---|
| $a$ (Å) | 3.8118(1) | 3.8106(1) | 3.8089(1) |
| $c$ (Å) | 6.1898(1) | 6.1606(1) | 6.1316(1) |
| Fe(1) - 2a | ¾ ¼ 0 | ¾ ¼ 0 | ¾ ¼ 0 |
| Fe(2) – 2c | ¼ ¼ 0.695(6) | ¼ ¼ 0.690(7) | ¼ ¼ 0.707(7) |
| $Ch$ – 2c | ¼ ¼ 0.2738(3) | ¼ ¼ 0.2719(3) | ¼ ¼ 0.2713(3) |
| Fe-$Ch$ bond length (Å) | 2.5504(1)×4 | 2.5369(1)×4 | 2.5287(1)×4 |
| Fe-Fe bond length (Å) | 2.6953(1)×4 | 2.6945(1)×4 | 2.6933(1)×4 |
| $Ch$-Fe-$Ch$ | 96.71(4)×2 | 97.36(4)×2 | 97.73(4)×2 |
| bond angle (deg) | 116.20(9)×4 | 115.85(9)×4 | 115.64(9)×4 |
| $R_F$ (%) | 2.22 | 2.53 | 3.01 |
| $R_B$ (%) | 3.28 | 3.55 | 3.65 |

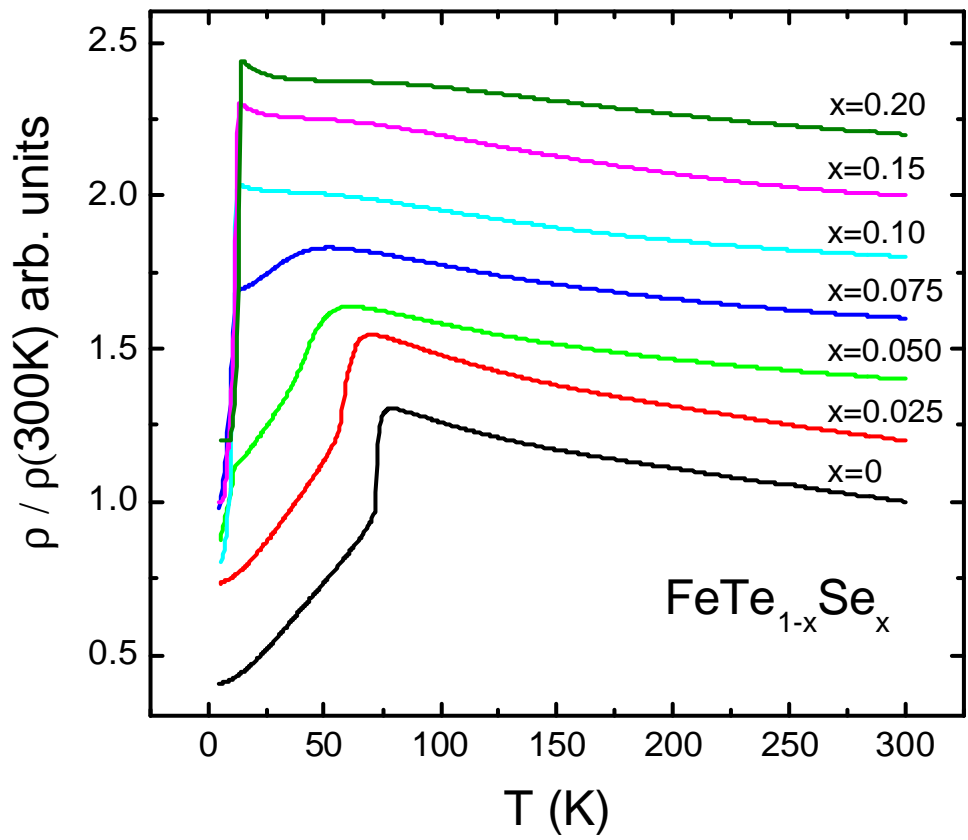

Figure 1 (color online): Resistivities for Fe(Te$_{1-x}$,Se$_x$) samples normalized to room temperature.

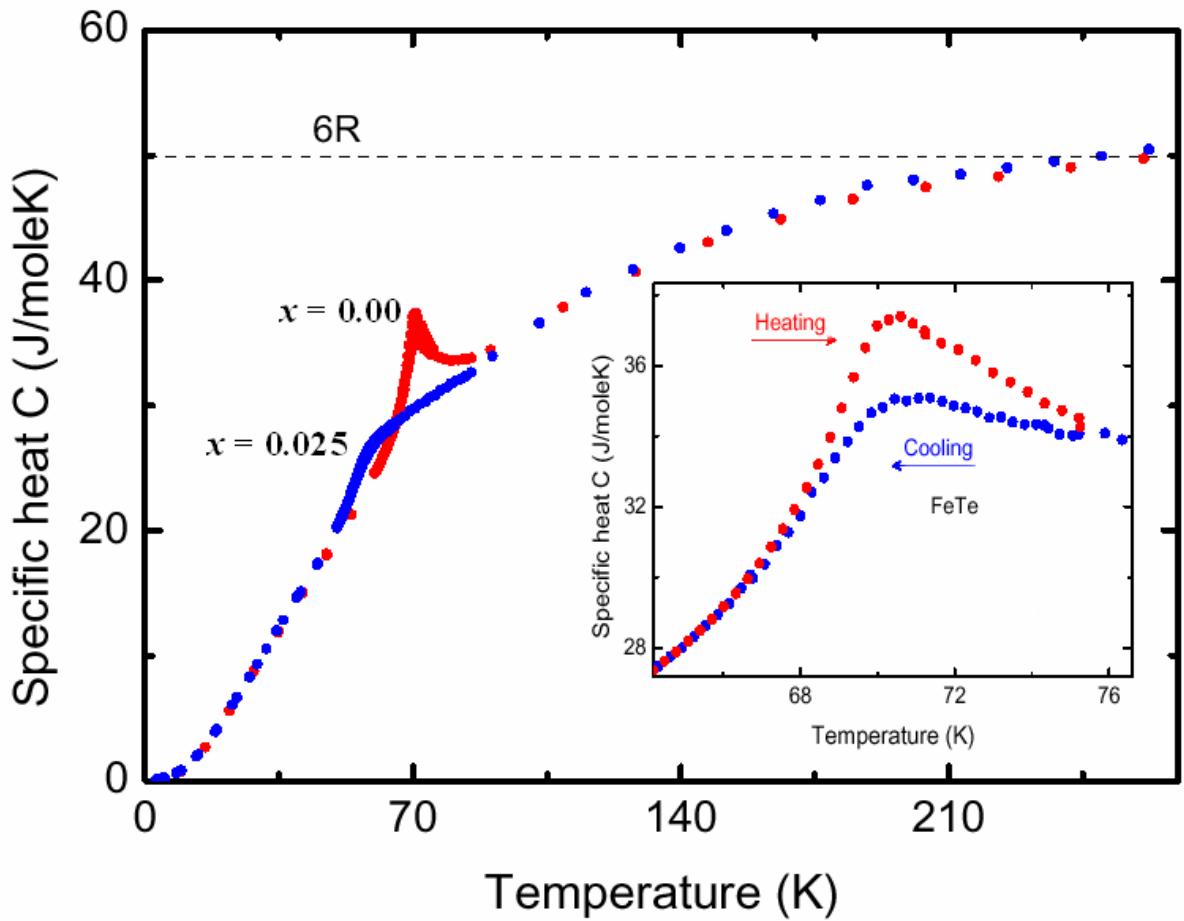

Figure 2 (color online): Temperature dependence of the heat capacity C(*T*) of FeTe and Fe(Te$_{0.975}$Se$_{0.025}$) samples; the inset shows the thermal hysteresis around transition for FeTe.

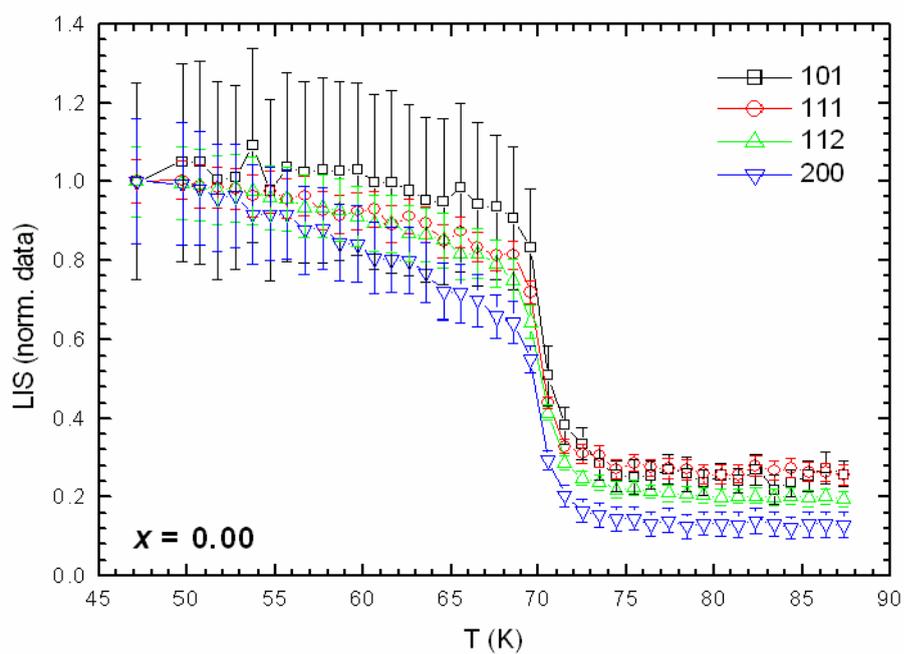

Figure 3 (color online): Evolution of the LIS with temperature (normalised data) for the 101, 111, 112 and 200 peaks of the FeTe sample.

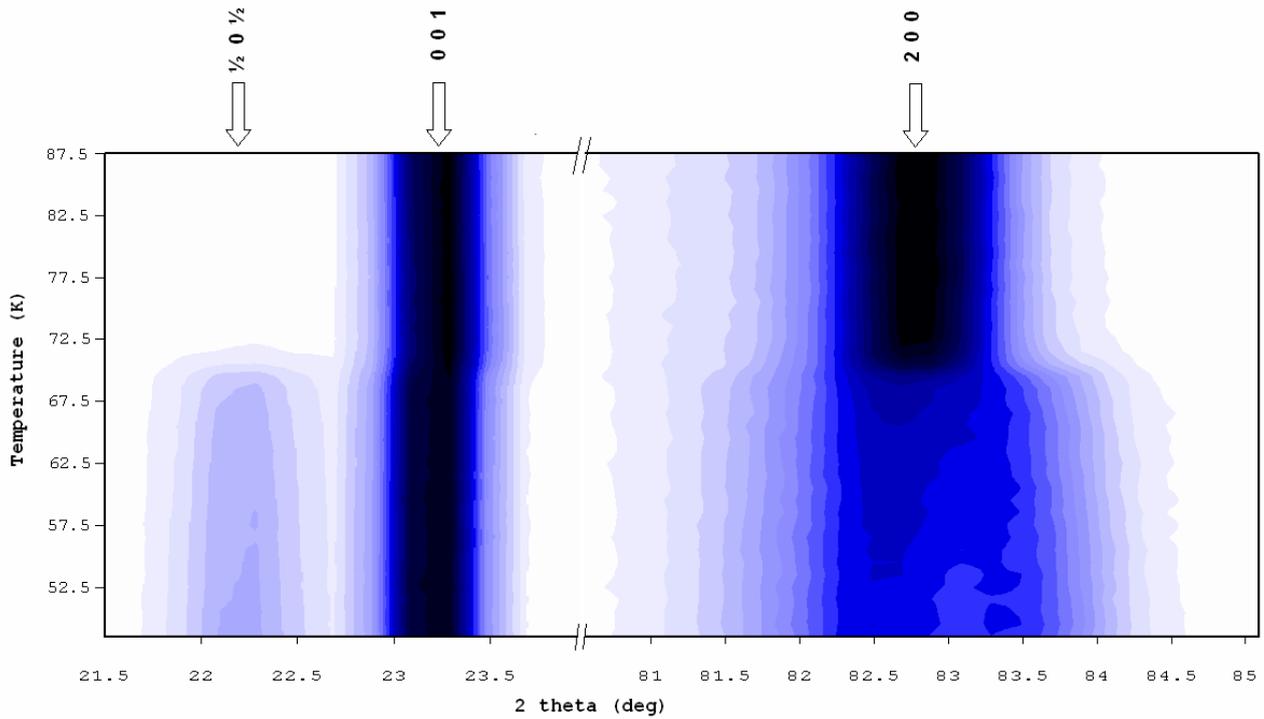
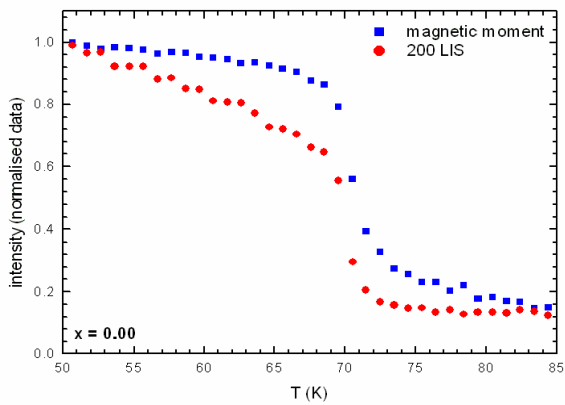
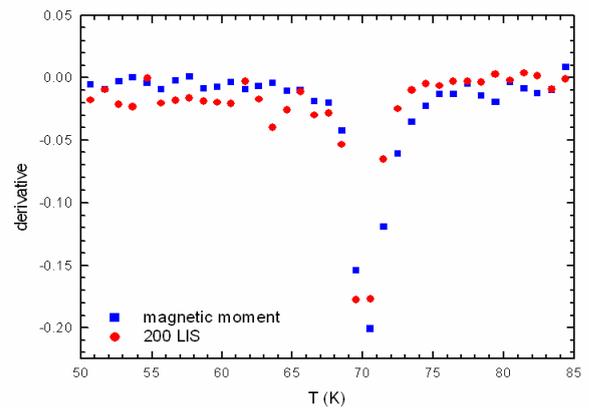

Figure 4 (color online): higher panel: thermodiffractogram of FeTe in the region of the ½0½ magnetic peak (plus 001 nuclear peak) and the tetragonal 200 nuclear peak, evidencing that magnetic ordering and structural transition occur at the same temperature (D1B data); lower panel: evolution of the ½0½ magnetic peak intensity and 200 nuclear peak LIS around the structural transition temperature (on the left) and their derivatives (on the right).

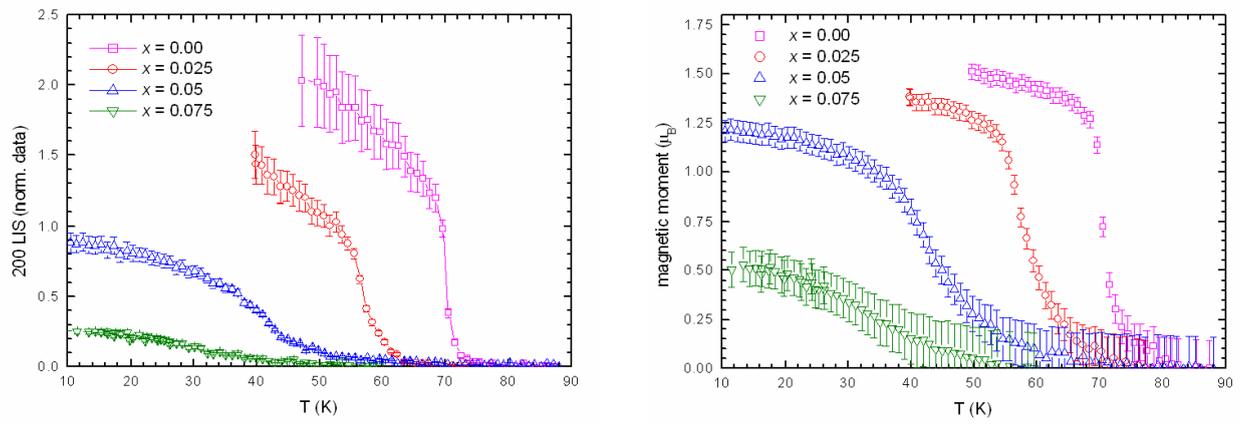

Figure 5 (color online): Evolution with temperature of the 200 peak LIS (on the right) and of the magnetic moment (on the left) for samples with $x \leq 0.075$.

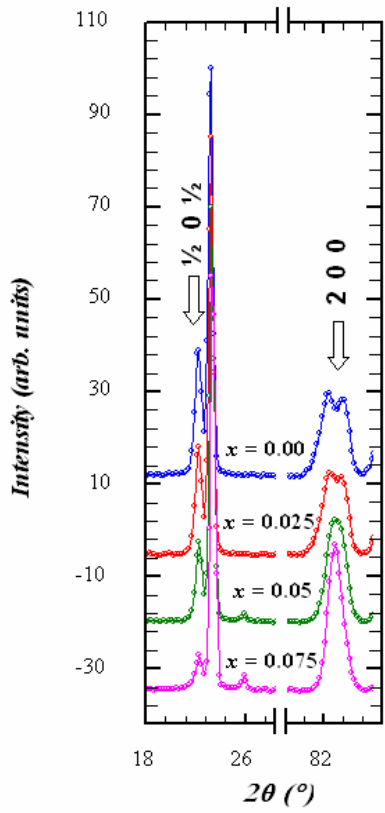

Figure 6 (color online): Comparison of the NPD patterns (D1B data) of the samples with $x \leq 0.075$ collected at 2 K on D1B in the regions of the ½0½ magnetic peak and the tetragonal 200 nuclear peak.

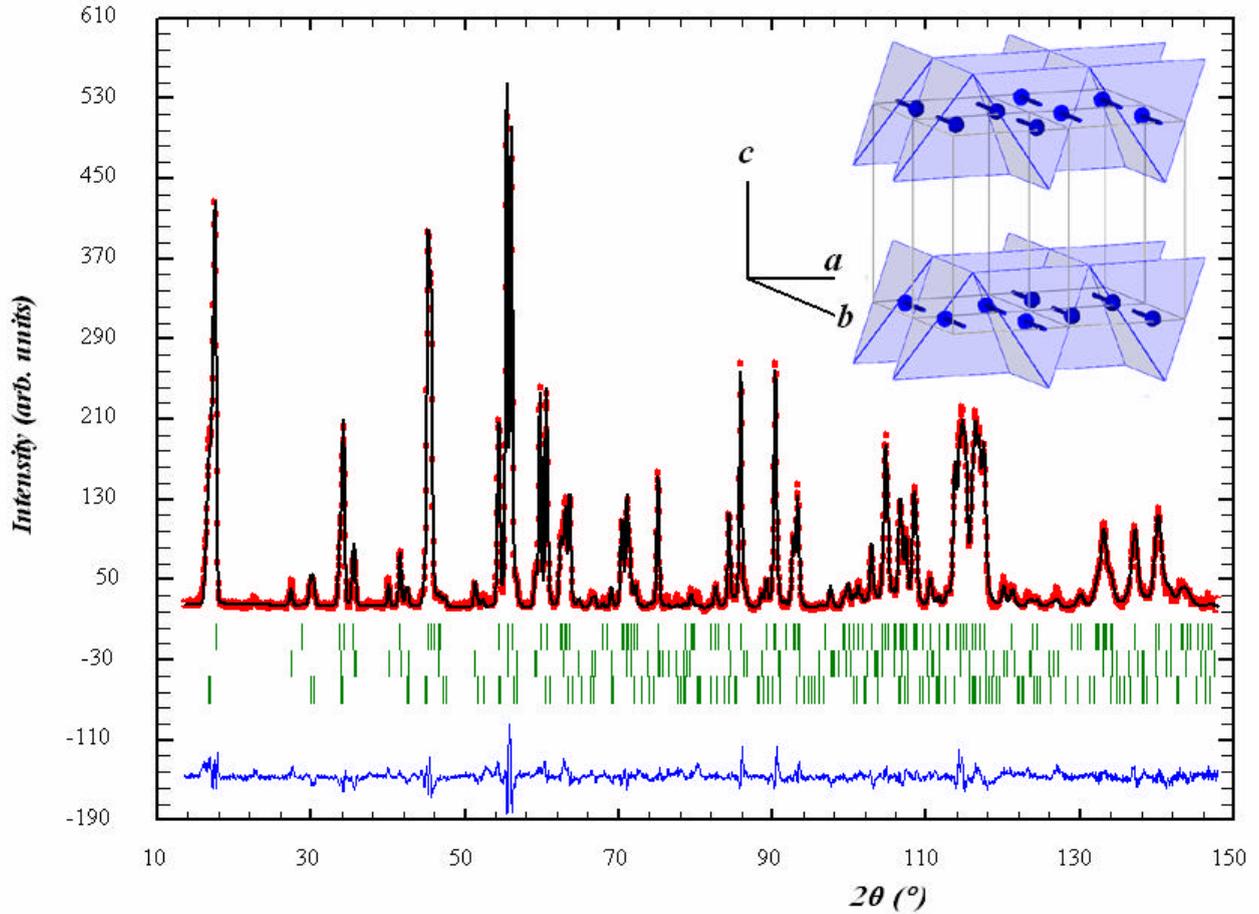

Figure 7 (color online): Rietveld refinement plot for FeTe (D1A data collected at 2 K). The points in the upper field represent the observed intensity data, the calculated pattern is superposed and drawn as a solid line; the small vertical bars indicate the position of the allowed Bragg reflections for the FeTe monoclinic nuclear phase (upper), the nuclear orthorhombic $FeTe_2$ phase (intermediate) and the Fe magnetic sub-lattice (lower); the difference between the observed and calculated patterns is plotted in the lower field. The inset shows the corresponding magnetic structure.

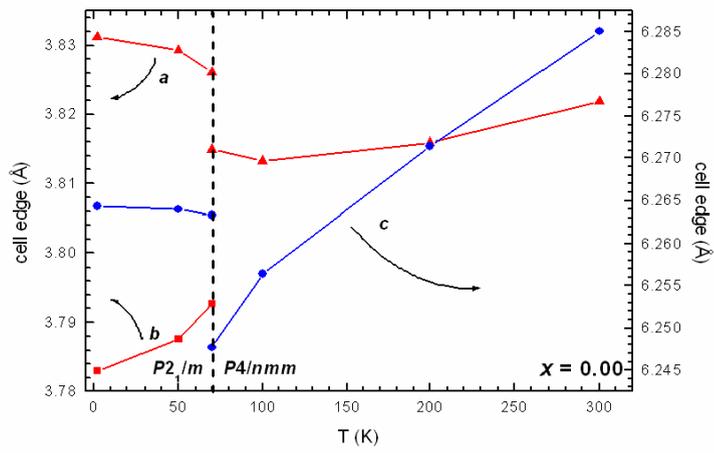

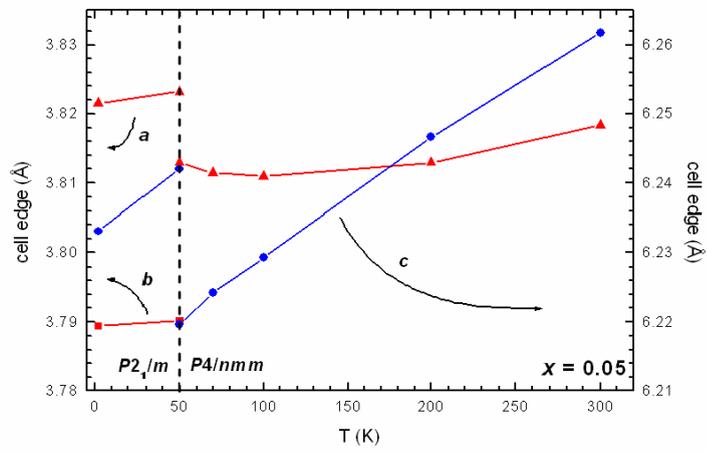

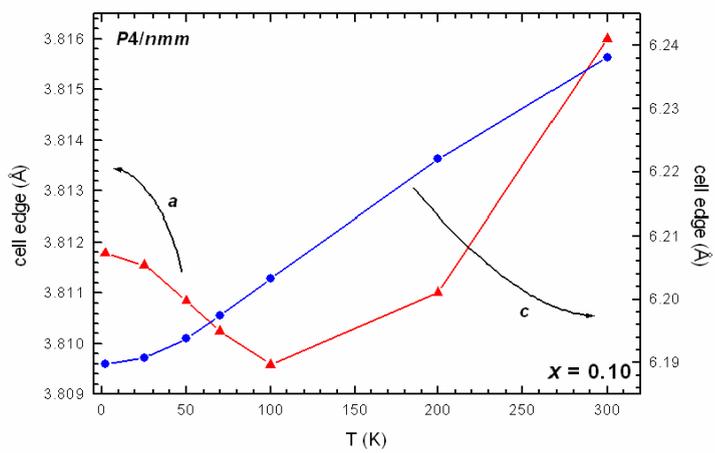

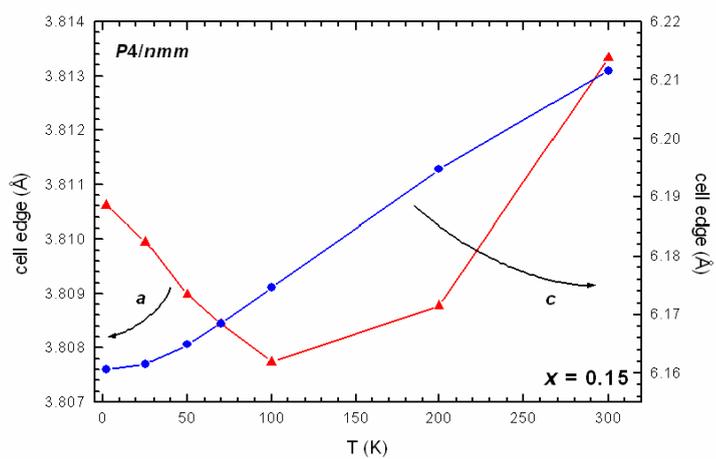

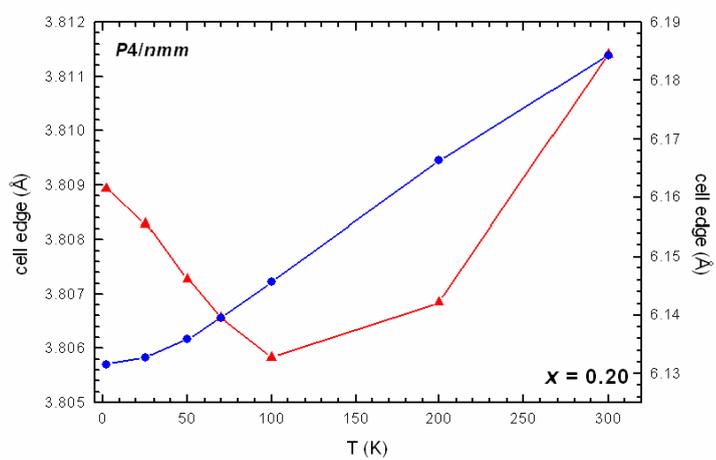

Figure 8 (color online): Evolution of the cell parameters as a function of $T$ below 300 K.

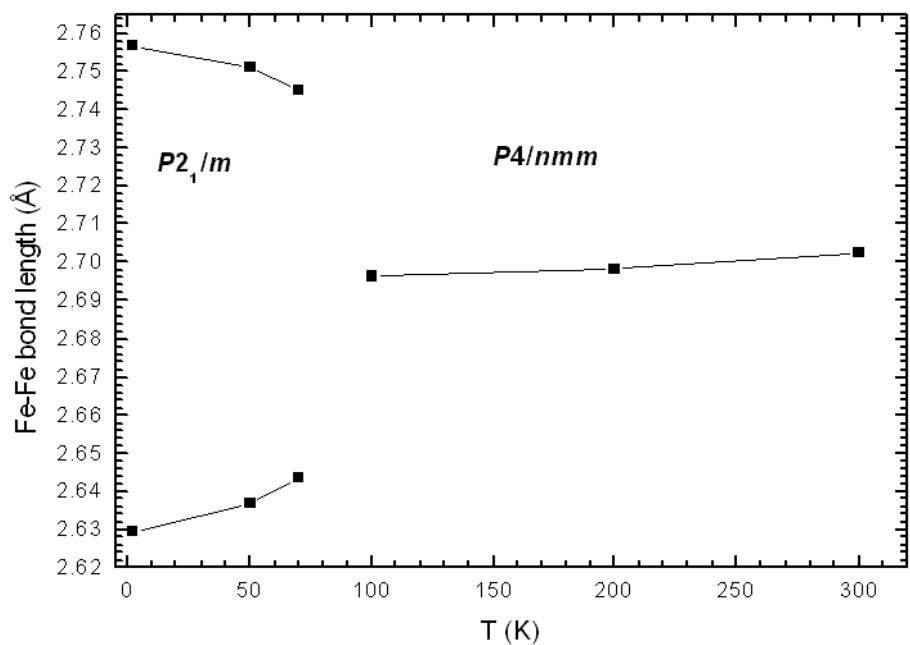

Figure 9: Evolution of the Fe-Fe bond lengths as a function of $T$ in FeTe.

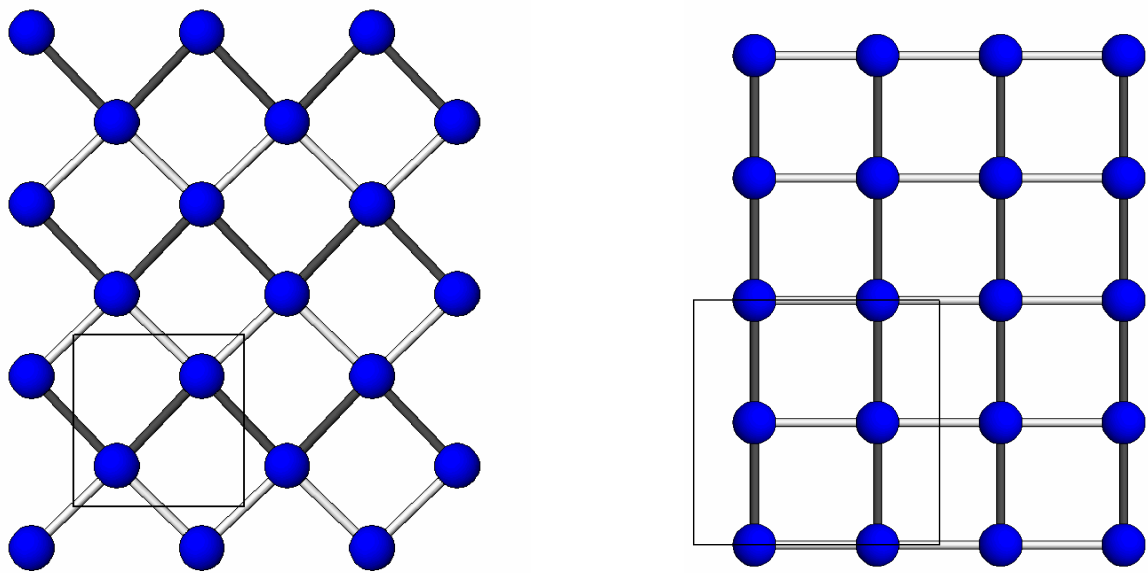

Figure 10: Stripe-like pattern originated by Fe-Fe bonds in monoclinic FeTe (on the left) compared to the pattern occurring in orthorhombic REFeAsO (on the right; shorter bonds: empty sticks, longer bonds: full sticks).

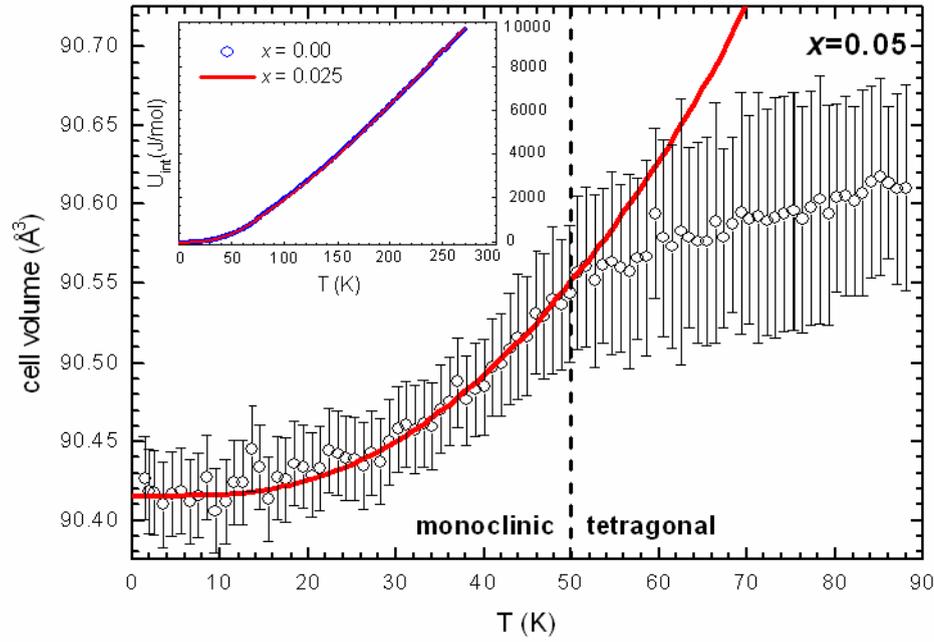

Figure 11: Evolution of the cell volume for *x* = 0.05 below 90 K (D1B data) fitted with Eq. 1 (solid curve); the inset shows the internal energy of the samples with *x* = 0.00 and 0.025 between 0 and 270 K, as evaluated by integrating specific heat data.

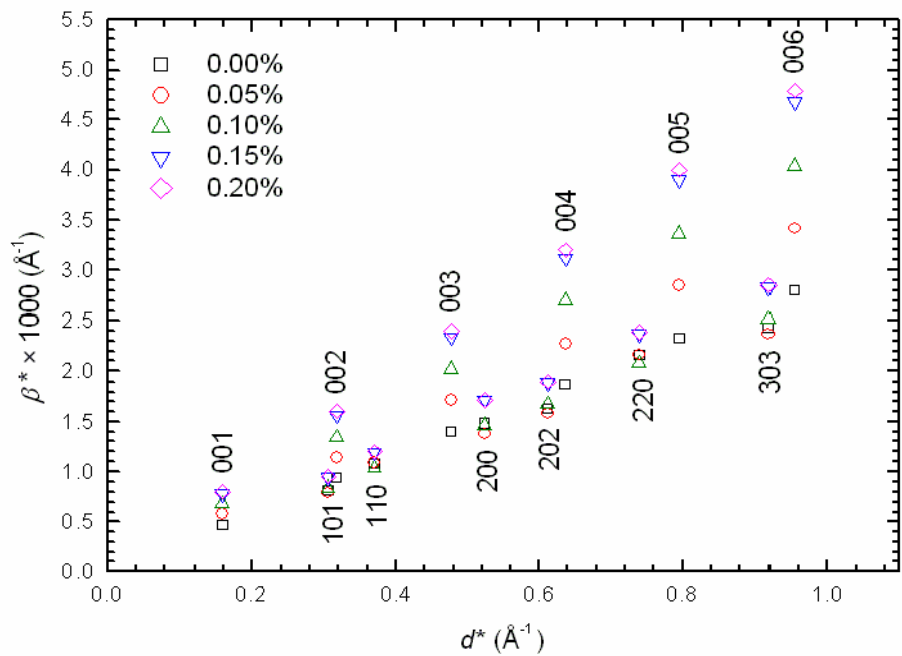

Figure 12 (color online): Superposition of the Williamson-Hall plots obtained by the microstructural analysis of the data collected at 300 K on D1A; only selected diffraction peaks are reported and indexed.

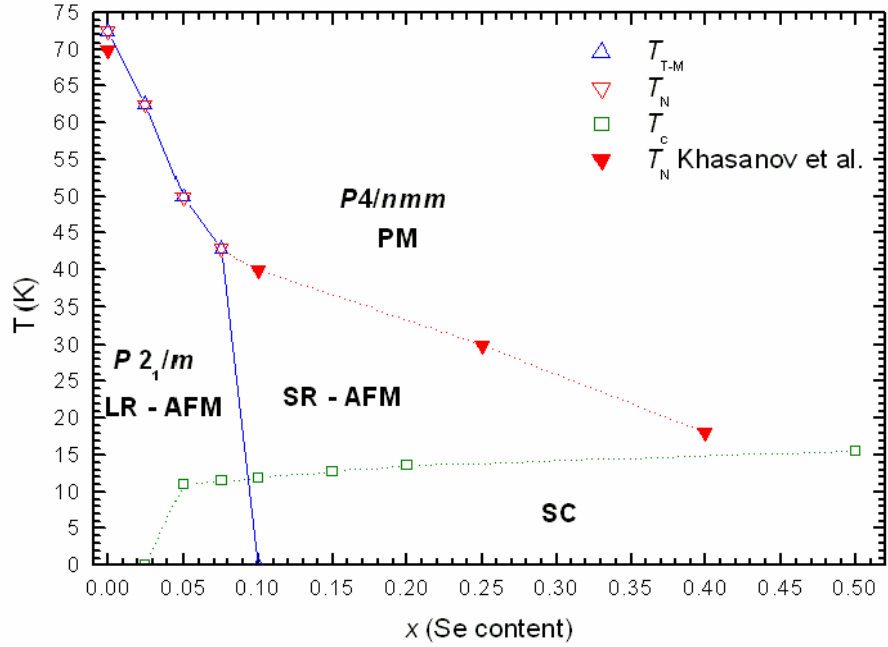

Figure 13 (color online): Phase diagram of the system Fe(Te$_{1-x}$,Se$_x$) for $x \leq 0.50$ showing the stability field of the LR-AFM interactions as obtained from our NPD data (empty symbols) compared with the field of the SR-AFM interactions reported by Khasanov *et al.*[8] (full symbols); note that the tetragonal to monoclinic phase transition is 1$^{st}$ order and hence, strictly speaking, the stability fields of the two polymorphs should be separated by a two-phase region.

# References

1 Kamihara, Y.; Watanabe, T.; Hirano, M.; Hosono, H., J. Am. Chem. Soc., 130 (2008) 3296

2 H. Okamoto, L. E. Tanner in *Binary Alloy Phase Diagrams*, vol 2, pg 1781, T. B. Massalski (ed.), ASM International (1990)

3 H. Okamoto in *Binary Alloy Phase Diagrams*, vol 2, pg 1769, T. B. Massalski (ed.), ASM International (1990)

4 T. M. McQueen, A. J. Williams, P. W. Stephens, J. Tao, Y. Zhu, V. Ksenofontov, F. Casper, C. Felser, R. J. Cava, Physical Review Letters 103 (2009) 057002

5 Wei Bao, Y. Qiu, Q. Huang, M. A. Green, P. Zajdel, M. R. Fitzsimmons, M. Zhernenkov, S. Chang, Minghu Fang, B. Qian, E. K. Vehstedt, Jinhu Yang, H. M. Pham, L. Spinu, Z. Q. Mao, Phys. Rev. Lett. 102, 247001 (2009)

6 Jinsheng Wen, Guangyong Xu, Zhijun Xu, Zhi Wei Lin, Qiang Li, W. Ratcliff, Genda Gu, J. M. Tranquada, Phys. Rev. 80, 104506 (2009)

7 Shiliang Li, C. de la Cruz, Q. Huang, Y. Chen, J. W. Lynn, Jiangping Hu, Yi-Lin Huang, Fong-Chi Hsu, Kuo-Wei Yeh, Maw-Kuen Wu, Pengcheng Dai, Phys. Rev. B 79, 054503 (2009)

8 R. Khasanov, M. Bendele, A. Amato, P. Babkevich, A. T. Boothroyd, A. Cervellino, K. Conder, S. N. Gvasaliya, H. Keller, H.-H. Klauss, H. Luetkens, V. Pomjakushin, E. Pomjakushina, B. Roessli, Phys. Rev. B 80, 140511(R) (2009)

9 J. Rodríguez–Carvajal, Physica B 192, 55 (1993)

10 M. Tropeano, I. Pallecchi, M. R. Cimberle, C. Ferdeghini, G. La mura, M. Vignolo, A. Martinelli, A. Palenzona, M. Putti, arXiv:0912.0395v1

11 D. K. Pratt, W. Tian, A. Kreyssig, J. L. Zarestky, S. Nandi, N. Ni, S. L. Bud'ko, P. C. Canfield, A. I. Goldman, R. J. McQueeney, Phys. Rev. Lett. 103, 087001 (2009)

12 S. Takeshita, R. Kadono, M. Hiraishi, M. Miyazaki, A. Koda, S. Matsuishi, H. Hosono, Phys. Rev. Lett. 103, 027002 (2009)

13 Y. Xiao, Y. Su, R. Mittal, T. Chatterji, T. Hansen, C. M. N. Kumar, S. Matsuishi, H. Hosono, Th. Brueckel, Phys. Rev. B 79, 060504(R) (2009)

14 C. Wang, Y. K. Li, Z. W. Zhu, S. Jiang, X. Lin, Y. K. Luo, S. Chi, L. J. Li, Z. Ren, M. He, H. Chen, Y. T. Wang, Q. Tao, G. H. Cao, Z. A. Xu, Phys. Rev. B 79, 054521 (2009)

15 A. Martinelli, A. Palenzona, M. Tropeano, C. Ferdeghini, M. R. Cimberle, C. Ritter, Phys. Rev. B 80, 214106 (2009)

16 A. Martinelli, M. Ferretti, P. Manfrinetti, A. Palenzona, M. Tropeano, M. R. Cimberle, C. Ferdeghini, R. Valle, C. Bernini, M. Putti, A. S. Siri, Supercond. Sci. Technol. 21 (2008) 095017

17 International Tables for Crystallography, vol. A, 5th edition T. Hahn (ed.), Springer (2005)

18 K. Horigane, H. Hiraka, K. Ohoyama, J. Phys. Soc. Jpn. 78 (2009) 074718

19 S. L. Bud'ko, P. C. Canfield, A. S. Sefat, B. C. Sales, M. A. McGuire, D. Mandrus, Phys. Rev. B 80, 134523 (2009)

20 J. I. Langford, D. Louër, E. J. Sonneveld, J. W. Visser, Powder Diffr. 1, 211 (1986)

21 E. Bellingeri, I. Pallecchi, R. Buzio, A. Gerbi, D. Marrè, M. R. Cimberle, M. Tropeano, M. Putti, A. Palenzona, C. Ferdeghini, arXiv:0912.0876v1

22 C.-Y. Moon, H. J. Choi, arXiv:0909.2916v1

23 A. M. Turner, F. Wang, A. Vishwanath, Phys. Rev. B 80 (2009) 224504

24 T. M. McQueen, A. J. Williams, P. W. Stephens, J. Tao, Y. Zhu, V. Ksenofontov, F. Casper, C. Felser, R. J. Cava, Phys. Rev. Lett. 103, 057002 (2009)